\documentclass[useAMS,usenatbib]{mn2e}
\usepackage{amsmath,amssymb,mathrsfs,graphicx,float,latexsym}
\usepackage{ragged2e}

\usepackage{xcolor}
\usepackage[colorlinks=true]{hyperref}
\hypersetup{citecolor=cyan,linkcolor=magenta}
\usepackage{multirow,array}
\usepackage{soul}
\usepackage[normalem]{ulem}
\newcommand{\md}{\rho_{\text{mean}}}
\newcommand{\Oo}{\Omega_{\text{orb}}}
\newcommand{\rot}{\Omega_{\text{s}}}
\newcommand{\cpt}{\mathcal{C}}
\newcommand{\tdm}{\tilde{\Lambda}}
\newcommand{\chirp}{\mathcal{M}}
\usepackage{journals}
\begin{document}
%%%%%%%%%%%%%%%%%%%%%%%%%%%%%%%%%%
% % %
\title[EOS groups from $g$-mode asteroseismology]{Constraining equation of state groups from $g$-mode asteroseismology}

\date{\today}
%%  Add Abbreviations for Journals

\author[Hao-Jui Kuan, Christian J.~Kr\"uger, Arthur G.~Suvorov, and Kostas D.~Kokkotas]{Hao-Jui Kuan$^{1,2}$,\thanks{E-mail:hao-jui.kuan@uni-tuebingen.de} Christian J.~Kr\"uger$^{1}$, Arthur G.~Suvorov$^{1,3}$, and Kostas D.~Kokkotas$^{1}$\\
	$^1$Theoretical Astrophysics, IAAT, University of T{\"u}bingen, T{\"u}bingen, D-72076, Germany\\
	$^2$Department of Physics, National Tsing Hua University, Hsinchu 300, Taiwan\\
	$^3$Manly Astrophysics, 15/41-42 East Esplanade, Manly, NSW 2095, Australia}

\date{Accepted XXX. Received YYY; in original form ZZZ}

\maketitle
\label{firstpage}
\pagerange{\pageref{firstpage}--\pageref{lastpage}} \pubyear{2021}

\begin{abstract}
\noindent{Buoyancy-restored modes inside neutron stars depend sensitively on both the microphysical (e.g., composition and entropy gradients) and macrophysical (e.g., stellar mass and radius) properties of the star. Asteroseismology efforts for $g$-modes are therefore particularly promising avenues for recovering information concerning the nuclear equation of state. In this work it is shown that the overall low-temperature $g$-space consists of multiple groups corresponding to different classes of equation of state (e.g., hadronic vs. hybrid). This is in contrast to the case of pressure-driven modes, for example, which tend to follow a universal relation regardless of microphysical considerations. Using a wide library of currently-viable equations of state, perturbations of static, stratified stars are calculated in general relativity to demonstrate in particular how $g$-space groupings can be classified according to the mean mass density, temperature, central speed of sound, and tidal deformability. Considering present and future observations regarding gravitational waves, accretion outbursts, quasi-periodic oscillations, and precursor flashes from gamma-ray bursts, it is shown how one might determine which group the $g$-modes belong to.}
\end{abstract}


\section{Introduction}

Neutron stars (NSs), with their extremely high core densities, reside in an otherwise hard-to-realise region of the QCD phase diagram. The details of the \text{(supra-)}nuclear equation of state (EOS) can only be accessed by connecting their observed outbursts or manifestations with, for example, theoretical predictions of the quasi-normal mode (QNM) spectrum, the maximal supportable mass, and gravitational radiation. Although substantial progress regarding constraints on the EOS from, e.g., the tidal deformability measured by the phase shift in the gravitational waveform for merging binaries \citep{LIGOScientific18,Raithel18,De18} and observations of the moments of inertia \citep{Greif20}, have been made in the literature, uncertainties remain and a wide pool of possibilities remain viable. Universal relations between certain stellar properties, however, offer additional avenues that are EOS-independent to infer inaccessible unknowns, and thus can indirectly narrow the space of valid candidate descriptions for the stellar interior.

In particular, the QNM spectrum of a NS is strongly associated with the global properties of the star, in the sense that several universal expressions relating mode frequencies and/or damping times to the bulk quantities of the star, like average density, moment of inertia, and tidal Love number, have been established for the $f$-, $p$-, and $w$-modes \citep{Andersson97,Lau09,Chan14,Kruger19,Benitez20,Sotani21}. This EOS-insensitive information provides hope for independent constraints by offering a critical tool in rephrasing the detected quantities in terms of others. For example, the mutual tidal deformability of a not-too-massive NSNS binary equates, in a roughly one-to-one fashion, into a compactness of the long-lived remnant because the $f$-mode properties of both the pre-merging and post-merger stars follow the same universal relation \citep{Manoharan21}. To account for the compositional structure, universal relations for $g$-modes, which distinguish one chemical configuration from another, need to be established.

In this article, we aim to introduce some global formulae for $g$-modes of NSs in various configurations and at various stages of their lives, either isolated or in binaries. The frequencies of this class of oscillatory modes are encoded in the Brunt-V{\"a}is{\"a}l{\"a} frequency, which measures the mismatch in the adiabatic index between the equilibrium and the mode-driven motions [see, e.g., \cite{Lai93}]. Using the (non-isentropic) general-relativistic pulsation equations, one of our main results is that we can encapsulate the Brunt-V{\"a}is{\"a}l{\"a} frequency into \emph{the temperature-modified mean density} of the star, and draw a relation to the $g$-mode frequency, viz.~
\begin{align}\label{eq:1}
	f_{g} \propto \md^{(1-\varsigma)/2}T,
\end{align}
for some EOS- and mode-number-dependent constant $-0.7\lesssim\varsigma\lesssim 0.5$ (see Sec. II). We also obtain two invertible mappings for the $g$-mode frequencies in terms of the \emph{central speed of sound}, $v_s$, and \emph{tidal deformability}, $\Lambda$, expressed as
\begin{align} \label{eq:rel2}
	\log\bigg(\frac{f_{g_1}}{100\text{ Hz}}M\bigg) =\sum_{i=0}^3 a_i(\log\Lambda)^{i},
\end{align}

\begin{align} \label{eq:rel3}
	\frac{f_{g_1}}{100 \text{ Hz}}\bigg(\frac{v_s}{R}\bigg) = \sum_{i=0}^3 b_i(\md)^{i/2},
\end{align}
with certain coefficients $\{a_i\}$ and $\{b_i\}$, stellar mass $M$, and radius $R$. Importantly, the first relation \eqref{eq:1} divides EOS into three quotient sets, and introduces the concept of \emph{$g$-space}, which characterises a specific EOS in terms of the $g$-mode spectrum. This occurs essentially because $\varsigma$ takes values in different blocks for different EOS families in a sense that is made precise in the main text. By representing a (hypothetically) observed $g$-mode as a point in $g$-space, we can, in principle, place constraints on the associated EOS for a wide range of systems including pulsars, progenitors of binary mergers, and the remnants of mergers. On the other hand, the latter two relations \eqref{eq:rel2} and \eqref{eq:rel3} are common for the set of twenty EOS considered in this article (i.e., the $a_{i}$ and $b_{i}$ are roughly constant among EOS). Leveraging this universality, we can deduce certain otherwise inaccessible quantities from $g$-mode observables. 

Although still a matter of debate, it has been suggested that various NS phenomena may be triggered by the excitation of $g$-modes. For instance, a small percentage of short gamma-ray bursts display precursor phenomena \citep{Coppin20,Wang20}, where energetic flashes are observed even many seconds prior to the main event in some cases. If the stellar oscillation modes briefly come into resonance with the orbital motion while the stars are inspiralling, significant amounts of tidal energy may be deposited into the mode(s), possibly to the point that the crust yields due to the exerted shear stresses exceeding the elastic maximum \citep{Tsang11,Tsang13}. $g$-modes in particular appear to lie in a sweet-spot, frequency wise, where the expected mode frequencies match the orbital frequency at the time of the precursor flashes \citep{Kuan21a,Kuan21b}. Similarly, $g$-mode frequencies match those observed in quasi-periodic oscillations (QPOs) of X-ray light curve from several NS systems and thus provide another promising avenue for detecting $g$-modes [see, e.g., \cite{Strohmayer13} for such speculation].

In addition, a new and quite important asteroseismological relation emerged through this study connecting the $f$-mode frequency $f_f$, with the radius and tidal deformability of an oscillating neutron star
\begin{align}
    \log\bigg(\frac{f_f}{\text{kHz}}R\bigg) 
    =\sum_{i=0}^2 c_i(\log\Lambda)^{i}\,.
\end{align}
This relation can be considered as complementary  to the one by \cite{Chan14}.

This paper is organised as follows. We introduce the grouping of EOS in terms of the range of $g$-mode frequencies, and we discuss the tidal effects of $g$-modes in Section \ref{sec.II}. We then turn to investigate the uniformity among the EOS set considered, producing some universal relations of $g$-modes in Section \ref{sec.III}. We speculate some circumstances, where we may be able to detect $g$-mode frequencies in Section \ref{sec.IV}. A discussion on the possible application of the obtained results is offered in Section \ref{sec.V}. Some EOS are found to allow for NSs with $g$-modes immune to external tidal field, which is discussed in Appendix \ref{appendix}.

Except where stated otherwise, we adopt the normalisation of $10 \text{ km} = 1 = \text{1 }M_{\odot}$ for the radius and mass of stars, and we reduce the dimension of velocity by unitising the speed of light, $c=1$.

\section{$g$-mode grouping}
\label{sec.II}
\begin{figure}
	\centering\hspace*{-.6cm}
	\includegraphics[scale=.4]{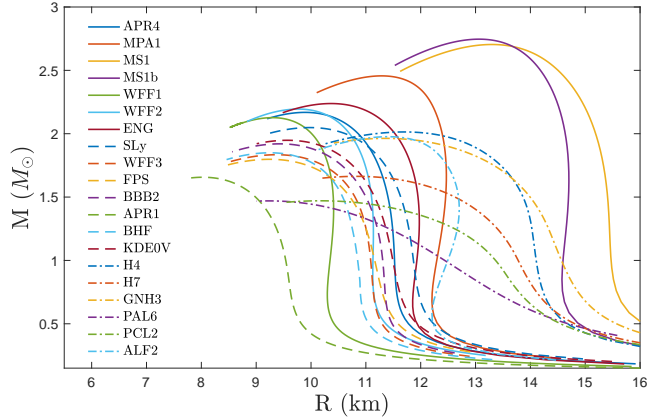}
	\caption{Mass-radius diagrams for the EOS considered here (see plot legends).}
	\label{fig:EOS}
\end{figure}
\begin{table*}
	\centering
	\caption{Fitting index $\varsigma$ defined in Eq.~\eqref{eq:fitomega2del} for a variety of EOS. Although there are a few outliers, we see that those listed in the first column have $\varsigma \gtrsim 0.2$, the second column features the range $0$-$0.2$, and in the final column are $\varsigma\lesssim0$.}
	\begin{tabular}{p{2cm}p{2cm}|p{2cm}p{2cm}|p{2cm}p{2cm}}
		\hline
		\hline
		\multicolumn{2}{c|}{Group I} & \multicolumn{2}{c|}{Group II} & 
		\multicolumn{2}{c}{Group III} \\
		\hline
		\hline
		EOS & $\varsigma$ & EOS & $\varsigma$ & EOS & $\varsigma$ \\
		\hline
		APR4$^{\text{a}}$ & 0.413 & SLy$^{\text{f}}$ & 0.158 & H4$^{\text{j}}$ & -0.094 \\
		\hline
		MPA1$^{\text{b}}$ & 0.321 & WFF3$^{\text{d}}$ & 0.027 & H7$^{\text{j}}$ & -0.221 \\
		\hline
		MS1$^{\text{c}}$ & 0.266 & FPS$^{\text{g}}$ & 0.011 & GNH3$^{\text{k}}$ & -0.081 \\
		\hline
		MS1b$^{\text{c}}$ & 0.333 & BBB2$^{\text{h}}$ & 0.101 & PAL6$^{\text{l}}$ & 0.059 \\
		\hline
		WFF1$^{\text{d}}$ & 0.453 & APR1$^{\text{a}}$ & 0.128 & PCL2$^{\text{m}}$ & -0.309 \\
		\hline
		WFF2$^{\text{d}}$ & 0.461 & BHF$^{\text{h}}$ & 0.143 & ALF2$^{\text{n}}$ & -0.657 \\
		\hline
		ENG$^{\text{e}}$ & 0.179 & KDE0V$^{\text{i}}$ & 0.200 &  &  \\
		\hline
	\end{tabular}
	\\
	\footnotesize{$\vphantom{\text{a}}^{\text{a}}$ \cite{APR4}.
	$\vphantom{\text{a}}^{\text{b}}$ \cite{MPA1}.
	$\vphantom{\text{a}}^{\text{c}}$ \cite{MS}.
	$\vphantom{\text{a}}^{\text{d}}$ \cite{WFF}.
	$\vphantom{\text{a}}^{\text{e}}$ \cite{ENG}.
	$\vphantom{\text{a}}^{\text{f}}$ \cite{SLy}.
	$\vphantom{\text{a}}^{\text{g}}$ \cite{FPS}.
	$\vphantom{\text{a}}^{\text{h}}$ \cite{BBB2}.
	$\vphantom{\text{a}}^{\text{i}}$ \cite{KDE0V}.
	$\vphantom{\text{a}}^{\text{j}}$ \cite{H}.
	$\vphantom{\text{a}}^{\text{k}}$ \cite{GNH3}.
	$\vphantom{\text{a}}^{\text{l}}$ \cite{PAL6}.
	$\vphantom{\text{a}}^{\text{m}}$ \cite{PCL2}.
	$\vphantom{\text{a}}^{\text{n}}$ \cite{ALF2}.}
	\label{tab:constant}
\end{table*}

We consider non-rotating, spherically symmetric equilibrium stellar models, obtained as solutions to the Tolman-Oppenheimer-Volkoff equations \citep{Tolman39,Oppenheimer39}, for several EOS approximated by piece-wise polytropic segments \citep{Read08,OBoyle20}; see Table \ref{tab:constant} for the EOS list, and Fig.~\ref{fig:EOS} for their mass-to-radius relations. The EOS are chosen to cover a wide range, from those obtained for NSs with purely $nep\mu$ matters, to those influenced by the hypothetically existing hyperons (e.g., GNH3, H4) possibly even displaying phase transition to free quarks (i.e., hybrid stars; e.g., ALF2). Although some of the considered EOS cannot support the heaviest pulsar observed to-date, namely PSR J0740+6620 \citep{NANOGrav19} with the mass of $\sim 2.14M_{\odot}$, they are included for the sake of demonstrating the generality of the trifurcation of EOS space, while we note that only EOS in Group I support the mass of PSR J0740+6620.

The polytropic (adiabatic) index of the star expresses the pressure exerted by a bulk of certain density, and depends on the detailed thermodynamic balance, chemical composition, and degeneracy status of the system \citep{Haensel02}. However, weak interactions (e.g., Urca processes) and/or diffusive processes within the star modulate the relative particle abundances over time \citep{Hoyos08,Hoyos10}, leading to entropy \citep{McDermott83,Reisenegger92} and/or compositional gradients \citep{Finn87}. These variations permit mobile fluid elements to experience buoyancy whenever perturbed away equilibrium, giving rise to the existence of $g$-modes. The frequencies of these modes therefore depend strongly on the strength of the stratification. Assuming that the adiabatic indices for the perturbation ($\Gamma$) and the equilibrium ($\gamma$) have a constant ratio $1+\delta$, we introduce a simple parameterisation for the forces supporting $g$-modes; see \cite{Kuan21a} for a discussion on the validity of this approximation. We identify $\delta$ with the temperature $T$ of isothermal stars in Sec.~\ref{sec.II.A}. Although we find that the correlation between $\delta$ and $T$ holds only for $T\lesssim 10^{10}$ K, we show the validity of the justification for binaries undergoing the tidal heating of $g$-modes Sec.~\ref{sec.II.B}. Respect to the simple parametrisation of stratification, three divisions of EOS are illustrated in Sec.~\ref{sec.II.C}. Overtones are denoted by $g_n$-mode, for which the radial eigenfunction has $n$ nodes.

\subsection{Temperature dependence}
\label{sec.II.A}

We consider EOS to be barotropic at the background level in the present article, hence $g$-modes are supported by the entropy gradient. In general, the temperature profile of NSs in different systems can differ significantly. For instance, mature, isolated NSs are expected to be almost isothermal \citep{Kruger14} [with the notable exception of magnetars, where the decaying magnetic field may provide an internal heat source \citep{Beloborodov16}], while NSs in binaries \citep{Perego19}, post-merger NS remnants \citep{Oechslin06,Kastaun16}, and especially accreting pulsars \citep{Potekhin03,Beloborodov16,Camelio20} can have spatially-dependent temperature profiles that span several orders of magnitude. The resultant position-dependent thermal pressure makes the ratio between the effective adiabatic index of the equilibrium and that for the perturbation a function of position.

For mature NSs, the temperature is typically low enough to allow the appearance of solid crust, superfluid core, and superconductivity of protons. Taking the multilayer complexity in full, $g$-modes can be distinguished into two classes, viz.~core and crust $g$-modes, depending on the configuration of the rendered motions \citep{McDermott85,McDermott88}. For instance, the radial motion caused by a core $g$-mode will be viscously-quenched at the crust-core interface, leading to only slight perturbations in the crust. The mode frequencies for both classes will acquire some modifications by the inclusion of superfluidity, which tends to increase the frequencies of crust $g$-modes. Nonetheless, we do not take the multilayer influences on the spectrum into account as the first step to reveal the trifurcation of EOS in terms of $g$-modes.

It has been shown that the $f$-mode frequency of a given star is largely determined by the square root of its mean density \citep{Andersson97,Kruger19},
\begin{align}
	\md = \frac{M}{\frac{4}{3}\pi R^3},
\end{align}
which is $\sim 0.33$ for typical values of mass and radius of NSs, viz.~$1.4M_{\odot}$ and $10$ km, in the units adopted here. We find a similar empirical relation but augmented with a dependence on stratification, given by 
\begin{align}\label{eq:fitomega2del}
	f_{g}/\sqrt{\delta} \propto \md^{(1-\varsigma)/2},
\end{align}
for quadrupolar ($l$=2) $g_1$-, $g_2$-, and $g_3$-modes in stars having mass larger than $1 M_{\odot}$ for a variety of EOS. Here $\varsigma$ is the fitting parameter, which varies for different EOS, while the proportionality constant depends on the mode quantum numbers for fixed overtone numbers ($g_1$-, $g_2$-, ...) as well as the EOS. However, despite this dependence, we find that it behaves in a quantitatively similar manner between different EOS groups (see Sec.~\ref{sec.II.C}). Furthermore, the fact that $g_1$- to $g_3$-modes follow the same relation indicates that the ratio between their frequencies can be approximated by a constant, e.g., we find that 
\begin{align}\label{eq:ratio12}
	f_{g_2}/f_{g_1}\simeq 0.68
\end{align} 
for every EOS considered. 

Assuming single component fluid NSs, the justification for $\delta$ introduced above may then be obtained by separating the contribution of the thermal pressure,
\begin{align}
	p_{\rm th}^{\rm x} = \frac{\pi^2}{6}\frac{n_{\rm x}}{E^{\rm x}_F}(kT)^2,
\end{align}
in the perturbation equations used to solve for QNMs, where ${\rm x} = \{\rm n,p,...\}$ runs over the different constituents of the fluid. Here $p^{\rm x}_{\rm th}$ represents the thermal pressure provided by the species ${\rm x}$, whose Fermi energy is $E^{\rm x}_F$. In particular, the thermal pressure contributes to the linearised equations of the radial and the tangential displacements [see~Eqs.~(B3) and (B4) of \cite{Kruger14}], which replace the term $\gamma p$ for NSs, where thermal gradients are absent, with [see~Eq.~(B5) of \cite{Kruger14}]
\begin{align}
	\Gamma p:=\gamma p+\sum_{\rm x}\frac{\partial p_{\rm th}^{\rm x}}{\partial n_{\rm x}}n_{\rm x} = \gamma(1+\delta) p,
\end{align}
where the temporally ($t$) and spatially ($\boldsymbol{x}$) varying parameter
\begin{equation}\label{eq:deltadef}
	\delta(t,\boldsymbol{x}) =\left[ \frac{k^2\pi^2}{6}\sum_x \frac{n_x(\boldsymbol{x})}{E_F^x(\boldsymbol{x})} \right] \frac{T(t,\boldsymbol{x})^2}{p(t,\boldsymbol{x})},
\end{equation}
is proportional to $T^2$. Although this implies that Eq.~\eqref{eq:fitomega2del} can be translated to
\begin{align}\label{eq:omega2T}
	f \propto \md^{(1-\varsigma)/2}T
\end{align}
if the NS is isothermal, such a parameterisation needs to be justified for an inhomogeneous temperature profile. To evaluate the reliability of Eq.~\eqref{eq:omega2T} for a NS with more general temperature distribution, we have compared the mode frequencies in isothermal stars with the cases  of stars with radial temperature profiles that fall by one to two orders of magnitude from centre to the surface by using the code described in \cite{Kruger14}. In particular, we solve for the frequencies of the $g_1$-mode for a particular stellar model with the following temperature profiles (in the unit of $K$): (i) isothermal with $\log T=\{10,9.5,9\}$; (ii) falling temperature by an order of magnitude as approaching the surface from the temperature at the center being $\log T=\{11,10\}$; (iii) falling temperature by more than one order of magnitude for a central temperature of $\log T=\{11,10.8,10.6,10.4,10.2,10\}$ and a fixed surface temperature $\log T=9$. The falling temperatures profiles are such that the temperature decreases linearly with the radius on a logarithmic temperature scale. As such, those profiles are highly artificial and given the multitude of different cooling and heating mechanisms operating inside a neutron star it would be difficult, and beyond the scope of the present work, to reproduce a realistic temperature profile. However, as we argue below, the most important part of the temperature profile is only that close to the surface of the star and since we are mostly interested in a qualitative understanding of temperature-related effects, such a simply constructed temperature distribution suits our needs.

We found that the spectrum of NSs with the same surface temperature differs only slightly among the above three scenarios, indicating that, to leading order, the QNM spectrum is unaffected by the temperature gradient except in extreme circumstances [cf.~the case of proto-neutron stars discussed by \cite{Torres-Forne19}; see also \cite{Sotani21b}]. This finding makes sense as the thermal pressure $p_{\rm th}$ is a function quadratic in the temperature; being causal for the buoyancy, it can compete with the static pressure, which grows substantially toward the center of the star, only in the outer regions of the star and, hence, the surface temperature is the primary quantity to determine the $g$-mode spectrum of a star. This argument also tells us that the shell below the surface which impacts the $g$-mode spectrum grows in depth with increasing surface temperature. We do observe this effect: Linearity between $f$ and $T$ was observed for $T\lesssim 10^{10}$ K, and a more complicated pattern arises only for $T>10^{10}$ K.

\subsection{Tidal Heating} 
\label{sec.II.B}

For stars within compact binaries permitting ``tidally-neutral $g$-modes'' (see Appendix \ref{appendix} for details), mode-related tidal effects are dominated by the non-resonant $f$-mode until merger. Resonant $g$-modes may contribute to the tidal effects comparable to $f$-mode for stars whose $g$-modes are tidally-susceptible, e.g., the produced stress may cause crust yielding \citep{Passamonti20} and the heating via shear viscosity may be as important as that resulting from the $f$-mode \citep{Lai93}. For the latter heating process, we find that the energy absorbed into the star via $g$-modes up to $g_{16}$-mode\footnote{We note that $\sim 80\%$ of the energy budget is deposited by $g_1$-$g_5$ modes. In addition, there is a caveat that high order $g$-modes, whose oscillatory period are smaller than the timescale of the involved reaction rate, are likely swept off from the spectrum \citep{Andersson19}.} 
can amount to $E_{\text{kin}}\sim 10^{43}$ erg for binaries consisting of identical components, where the kinetic energy of each mode acquired during the tidal excitation is consistent within a factor of $\sim 5$ with the stationary-phase-approximated formula [see Eq.~(6.11) in \cite{Lai93}],
\begin{align}
	E_{nl} & \simeq 9.70\times10^{42} \text{erg}\nonumber\\
	&\times\bigg(\frac{\md}{0.49}\bigg)^{-1}
	\bigg(\frac{f_{nl}}{100\text{ Hz}}\bigg)^{1/3}
	\bigg(\frac{Q_{ln}}{10^{-5}}\bigg)^2.
\end{align} 
A portion of the mechanical energy will be converted into thermal energy via viscosity \citep{Lai93,Sherf21} mainly provided by lepton shear viscosity when $T< 10^9$ K [cf.~Fig.~(1) of \cite{Andersson04} and also \cite{Kolomeitsev14}], and chemical reactions, such as direct or modified Urca processes \citep{Arras18}. To set an upper limit for the heating effect, we assume the whole energy budget deposited by the finite series of $g$-modes is converted into thermal energy through the aforementioned channels; consequently, NSs will be heated up to a temperature 
\begin{align}
	T \simeq 4.7\times10^6\bigg(\frac{E_{\text{kin}}}{10^{43} \text{ erg}}\bigg)^{1/2} \text{ K},
	\label{eq:temp}
\end{align}
by equating the dissipated energy to the heat content $U$ of NSs \citep{Lai93,Arras18}, which related to the averaged temperature over the core via [Eq.~(8.28) in \cite{Lai93}]
\begin{align}
	U \simeq 4.5\times10^{45} \bigg(\frac{T}{10^8 \text{ K}}\bigg)^2 \text{erg}.
\end{align}
Although $f$-mode excitation may also extract orbital energy thus attributing to the tidal heating, the amount of the kinetic energy is estimated to be comparable to the gross value of $g$-modes \citep{Lai93}. The square root dependence of the temperature to the kinetic energy, as expressed in Eq.~\eqref{eq:temp}, therefore indicates the resultant temperature will be of the same order, which is cold enough not only for the appropriateness of our implementation of cold EOS, but for the justification of $\delta\propto T^2$.

We note that the above temperature may be too cold to admit $g_1$-mode with frequency of $\gtrsim 100$ Hz since we estimate the heating effect for non-spinning NSs. For the rotating stars, $r$-modes, even $f$-mode, may undergo a period of resonance \citep{Ho98,Lai06} thus soaking more energy in expense of orbital one to heat further the star.

\subsection{Three Groups}
\label{sec.II.C}
\begin{figure}
	\centering\hspace*{-.6cm}
	\includegraphics[scale=0.35]{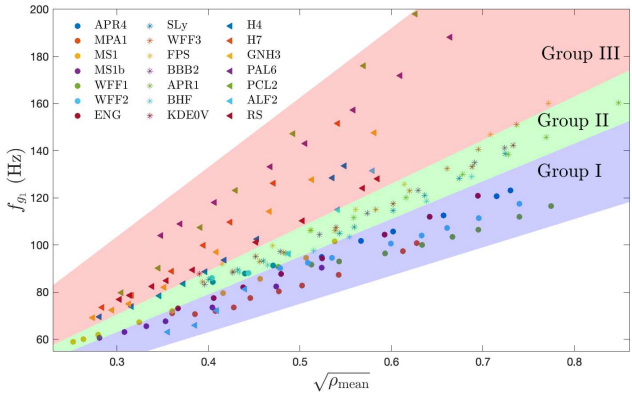}
	\caption{$g$-mode frequency for $\delta=0.005$ as a function of the square root of the mean density, $\sqrt{\md}$, for each of the EOS under study. The circles represent those models drawn with solid line in Fig.~\ref{fig:EOS}, the asterisks pair the dashed-line EOS, and the triangles correspond to the dash-dot EOS.}
	\label{fig:grouping}
\end{figure}
\begin{figure}
	\centering\hspace*{-5mm}
	\includegraphics[scale=0.46]{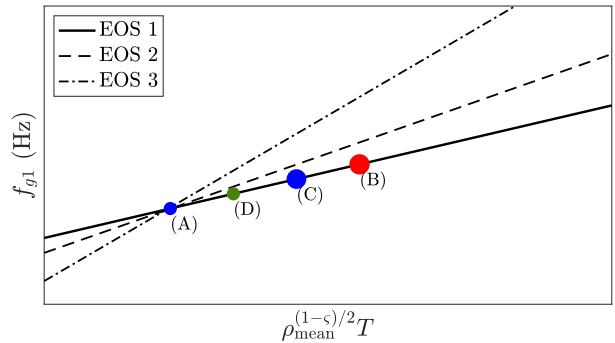}
	\caption{Symbolic \emph{$g$-space} (see main text for the definition) for three EOS -- one from each group, and four hypothetical observations. The chosen EOS trifurcates around a certain point in the space, so that all three branches can explain observation (A). However, suppose that, for demonstration purposes, Group I (EOS 1) governs NS from different systems (and thus passes through each observation). The member(s) in Group II (EOS 2) can fit the observation in system (A), and may be able to explain observation (D) depending on the error bars of the measurement, though are incapable of matching the data for (C) or (B). The group III case (EOS 3), by contrast, is unable to explain any observation except (A).}
	\label{fig:g_space}
\end{figure}

In Eq.~\eqref{eq:fitomega2del}, the exponent is determined as follows: for a certain value of $\varsigma$, a least-squares fitting is applied to $f_g/\sqrt{\delta}$ and $\md^{(1-\varsigma)/2}$, from which we calculate the correlation coefficient; we then define the parameter to be the value of $\varsigma$ that maximises the correlation coefficient. We collate the fitting parameters $\varsigma$ in Tab.~\ref{tab:constant}. We see that $\varsigma$ can be divided into three ranges (with a few outliers): $\gtrsim 0.2$ (\emph{Group I}; first column), $\sim0$--$0.2$ (\emph{Group II}; second column), and $\lesssim 0$ (\emph{Group III}; final column). The width of the 95\% confidence interval of $\varsigma$ for each EOS is $\lesssim0.01$, indicating the grouping presented here is statistically sharp. The three divisions intriguingly correspond to EOS consisting of only hadronic matters that can support heavy stars of mass $\gtrsim2.12M_{\odot}$ or more (Group I), those that cannot (Group II; $\lesssim2.07M_{\odot}$ or less), and EOS involving phase transitions leading to either hyperon condensation or quark deconfinement (Group III). We note that the considered EOS of Group III cannot support stars much heavier than $\sim2M_{\odot}$. The few two peculiar EOS are ENG and PAL6; the first is an outlier of \emph{Group I} with $\varsigma<0.2$, while the latter EOS is an exception of \emph{Group III} since it has a positive $\varsigma$ and the matter consists of hadrons. This slight mixing between Group III and the others is not entirely surprising, since first-order transitions in EOS that only support relatively light stars will not play a big role. In general, we see that the mass-radius relation is `flattened' with decreasing $\varsigma$, and becomes a one-to-one mapping when $\varsigma\lesssim 0.2$ except for the Group I EOS ENG and the Group III EOS ALF2. A natural question about this grouping is whether they are classified by ``softness''? While this may be a general trend it is not absolute since, for example, the stiffest EOS are MS1 and MS1b, which are in Group I, while members in Group III tend to be stiffer than those in the other two groups. 

We illustrate the nature of the groups by plotting the $g$-mode frequency for a fixed $\delta$ as a function of (the square root of) $\md$ in Fig.~\ref{fig:grouping}, where the different markers are used to indicate the different correlations between the range of $\varsigma$. For instance, EOS for which $\varsigma\gtrsim 0.2$ (group I) are represented by solid circles. One observes that the $g$-mode frequencies even of very stiff EOS like MS1 and MS1b align well [in the sense of Eq.~\eqref{eq:fitomega2del}] with other EOS of Group I, especially when the difference in the radius for NSs with a fixed mass among these EOS can be as large as $1.5$ times (between WFF1 and MS1). This observation is similar for respective members of the other two groups, clearly indicating that global parameters associated with the EOS are not the only factor in contributing to $g$-modes. On the other hand, it has been shown that $g$-modes in proto-NS following a core collapse, which may be detected in the near future by third generation detectors \citep{Ott06,Bizouard20}, tend to obey a universal relation, and there is no obvious partition [see, e.g., Fig.~(2) in \cite{Torres-Forne19}]. The absence of grouping in proto-NSs may arise for two reasons: (i) they have very high temperature,  likely indicating that they are thermally-stratified to the degree that $\delta\gtrsim1$ (for which the correlation we find between $T$ and $\delta$ may no longer hold), and thus the situation may be different from cold NSs (see Sec.~\ref{sec.II.C} for the discussion on valid range of the introduced parameter $\delta$); (ii) the EOS families that have been previously considered in this context may incidentally belong to the same branch. Although beyond the scope of this paper, it would be worth revisiting $g$-mode studies in proto-NS with a wider EOS library to investigate these points.

Cold NSs with similar bulk properties display universal relations between the $f$-, $p$- and $w$-modes, indicating the particulars of internal structure does not impact these modes much \citep{Andersson97,Kruger19,Benitez20,Sotani21}. However, the compositional content of NSs, which is to a large extent unknown, will substantially affect the $g$-mode spectrum. The uncertainty in the internal physics, on top of the various methods and/or assumptions adopted to model the nuclear interactions, leads to the richness in variations of EOS candidates; the grouping we observe therefore indicates that microphysical considerations can be broadly categorised into three channels, each leading to a family of $g$-modes in Fig.~\ref{fig:grouping}.  For instance, the constituents in the core depend on their evolution track since certain cooling/reaction channels have density-dependent activation thresholds. In addition, atomic abundances in the crust may vary from system to system, depending on their birth site \citep{Leszczynska15,Woosley20}. Crustal variations are unlikely to considerably modify the \emph{core} $g$-mode spectrum however, and thus would not be responsible for the grouping. 

Due to the limited precision of any given observation, it is likely that we can only distinguish one family of EOS from another rather than two EOS in the same family. If it happens that there are two phenomena caused by $g$-modes, and these are found to reside in different groups, it might imply that the NSs have followed a different evolutionary track as far as the EOS is concerned. Nonetheless, if we assume that a certain functional EOS applies to NSs in different systems, such as binaries and long-lived remnants of mergers, these compact objects will belong to the same line (solid line in Fig.~\ref{fig:g_space}) describing the $g$-mode frequency as a function of $\md^{(1-\varsigma)/2}T$, just at different stages. Defining the $g$-space as the set consisting of the $g_{n}$-mode frequencies that can be possessed by at least one star with this EOS, we indicatively draw the $g_1$-space in Fig.~\ref{fig:g_space} for EOS 1-3, each belongs to Group I-III, respectively.  The observations of several events can therefore be incorporated to provide a mutual constraint on the EOS candidates (see below) but in a manner that is distinct from Bayesian analyses, where various observations shape the prior of the parameterisation of EOS differently to gradually reduce the viable region for EOS candidates [see, e.g., \cite{Miller19,Raaijmakers19,Raaijmakers21}]. We note that in a formal, Bayesian analysis, this common EOS assumption will shape the (informative) prior of model parameters differently than the situation when the assumption is absent, thus affecting the statistical inference \citep{Kastaun16}.

As an illustrative example, we consider the particular combination of systems; (A) represents a NS in a binary undergoing the inspiral, (B) stands for a newly-born NS from merger after the temperature has dropped sufficiently to validate Eq.~\eqref{eq:omega2T} and differential rotation has largely stabilised
%
%for Shakura-Sunyaev viscosity with $\alpha_{\text{SS}} \sim 10^{-2}$ 
\footnote{A NS remnant from a merger forms in a differentially-rotating state, and is likely to be highly magnetised due to the Kelvin-Helmholtz instability occurring at the shear boundary formed upon the contact of the two progenitors \citep{Kiuchi15}. The non-uniform rotation will wind the field lines and is thought to produce a turbulent viscosity however, which diffuses the angular momentum from the fast to slow regions, thus unifying the rotation profile (or at least confining it to vary only along flux lines in accord with the axisymmetric Ferraro theorem) in just a few tens of milliseconds \citep{Duez04,Fujibayashi17}.}, 
and (C) denotes an old, long-lived remnant from an NS merger, with finally a less-compact version in (D) (see also below). Each type of observation from different NSs is designated as a point in the space spanned by $f_{g_1}$ and $\md^{(1-\varsigma)/2}T$, and the solid line connecting these three points indicates a hypothetical case where one EOS ``branch'' satisfies (A)-(D). We see that the lines for different groups diverge, but with only one measurement point (A), one cannot tell which line it is unless some additional observations are available; branches can be ruled out by incorporating the multi-stage information across (A) to (C), depending on the relative error bars on the measurements. Nonetheless, we emphasise that the above analysis surrounding Fig.~\ref{fig:g_space} does not necessarily reflect the reality. Instead, it is simply a demonstration of the idea.

In order to map a certain system onto the $g$-space, the mean density, which implies we are able to probe the mass and radius to certain extent, and volume-averaged temperature, must be known. Although certain constraints can already be set by the simultaneous determinations of the mass and the radius by, e.g., marking a valid region on the mass-radius diagram for EOS, there will still be a bunch of EOS surviving such restriction if the radius has even small error bars \cite[e.g.,][]{Lindblom92,Weih:2019rzo}. Information from the $g$-mode frequency can, in such cases, provide additional constraints of a different flavour. Considering a NS with a canonical mass of $1.5M_{\odot}$ and a radius of $11-12$ km (corresponding to $\sqrt{\rho_{\text{mean}}}=0.46-0.52$), for example, several members of Group I and II are adequate in terms of the mass and the radius observations, viz.~APR4, MPA1, WFF2, SLy, WFF3, FPS, KDE0V, and BBB2 \cite[though considering higher mass or more compact stars reduces the pool;][]{Weih:2019rzo}. Some of these EOS will be ruled out if $f_{g_1}$ can be acquired; for instance, $f_{g_1}\lesssim80$ Hz will exclude ones belonging to Group II thus making a 62.5\% reduction in the aforementioned EOS candidates (see.~Fig.~\ref{fig:grouping}).

To observe a given system on the $g_1$-space, there must be some phenomena attributed to its $g_1$-mode. The candidate systems allowing for $g$-mode measurement will be discussed in Sec.~\ref{sec.IV}, where we consider electromagnetic precursor flares prior to merger (Sec.~\ref{sec.IV.A}), and quasi-periodic oscillations (QPOs) in the X-ray light curves of accreting millisecond X-ray pulsars (AMXPs; Sec.~\ref{sec.IV.B}).

\section{Universal relations}
\label{sec.III}

In addition to the EOS-dependent relation \eqref{eq:fitomega2del} [or Eq.~\eqref{eq:omega2T}], we provide three EOS insensitive relations bearing $g$-mode frequencies to $f$-mode frequency, dimensionless tidal deformability $\Lambda$ (Sec.~\ref{sec.III.A}), and the central speed of sound (Sec.~\ref{sec.III.B}), respectively. These relations offer extra means not only to distinguish EOS groups, but to extract information on the properties of NSs from gravitational- and/or electromagnetic-radiation observations. We consider hereafter only EOS of Group I and II, pondering that these are favoured by precursor events (see the discussion near the end of Sec.~\ref{sec.IV.A}). 

\subsection{Gravitational Waves}
\label{sec.III.A}
When analysing a high signal-to-noise gravitational waveform, the chirp mass,
\begin{align}
	\chirp = \frac{(M_1M_2)^{3/5}}{(M_1+M_2)^{1/5}},
\end{align}
 of the system can be determined to a relatively high precision since it enters the phase evolution at the lowest post-Newtonian order \citep{Peters63,Cutler94}. Although not with the same level of accuracy, the symmetric mass ratio,
\begin{align}
	\eta = \frac{M_1M_2}{(M_1+M_2)^2},
\end{align}
can also be obtained by exploiting its post-Newtonian order contribution to the GW phase [see, e.g., Eq.~(32) in \cite{Kokkotas94}]. Therefore, the individual masses $M_1$ and $M_2$ can be estimated from $\chirp$ and $\eta$. In addition, a detailed extraction of the phase shift $\delta\Psi=\delta\Psi_{\text{eq}}+\delta\Psi_{\text{dyn}}$ may shed light on both equilibrium and dynamical tidal effects \citep{Pratten19}, for which the mutual, dimensional tidal deformability \citep{Hinderer09,LIGOScientific17},
\begin{align}\label{eq:deforma}
	\tdm = \frac{16}{13}\frac{(M_{1}+12M_{2})M_{1}^{4}\Lambda_1 +(M_2+12M_1)M_{2}^{4}\Lambda_2}{(M_1+M_2)^5},
\end{align}
is the agency of the former, and the latter is mainly produced by the late stage growth of $f$-mode \citep{Hinderer16,Schmidt19,Steinhoff21}. Here $\Lambda_{1(2)}$ is the tidal deformability of the primary (companion). With observables $\chirp$, $\eta$, and $\tdm$, we get a relation for $\Lambda_1$ and $\Lambda_2$, and the masses $M_1$ and $M_2$ though with certain error. However, the uncertainty of EOS involved in such analysis has a ramification on measuring stellar properties, which can be mitigated by implementing EOS-insensitive relations. 

\begin{figure}
	\centering\hspace*{-.6cm}
	\includegraphics[scale=.4]{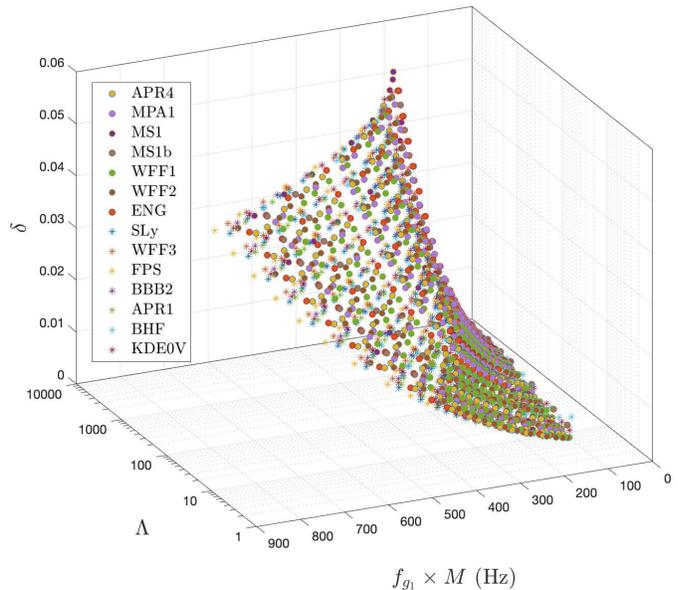}
	\caption{Three-dimensional universal relation among mass-scaled $g_1$-mode frequency, the tidal deformability, and the stratification for EOS in Group I and II.}
	\label{fig:sheet}
\end{figure}
\begin{figure}
	\centering
	\includegraphics[scale=0.43]{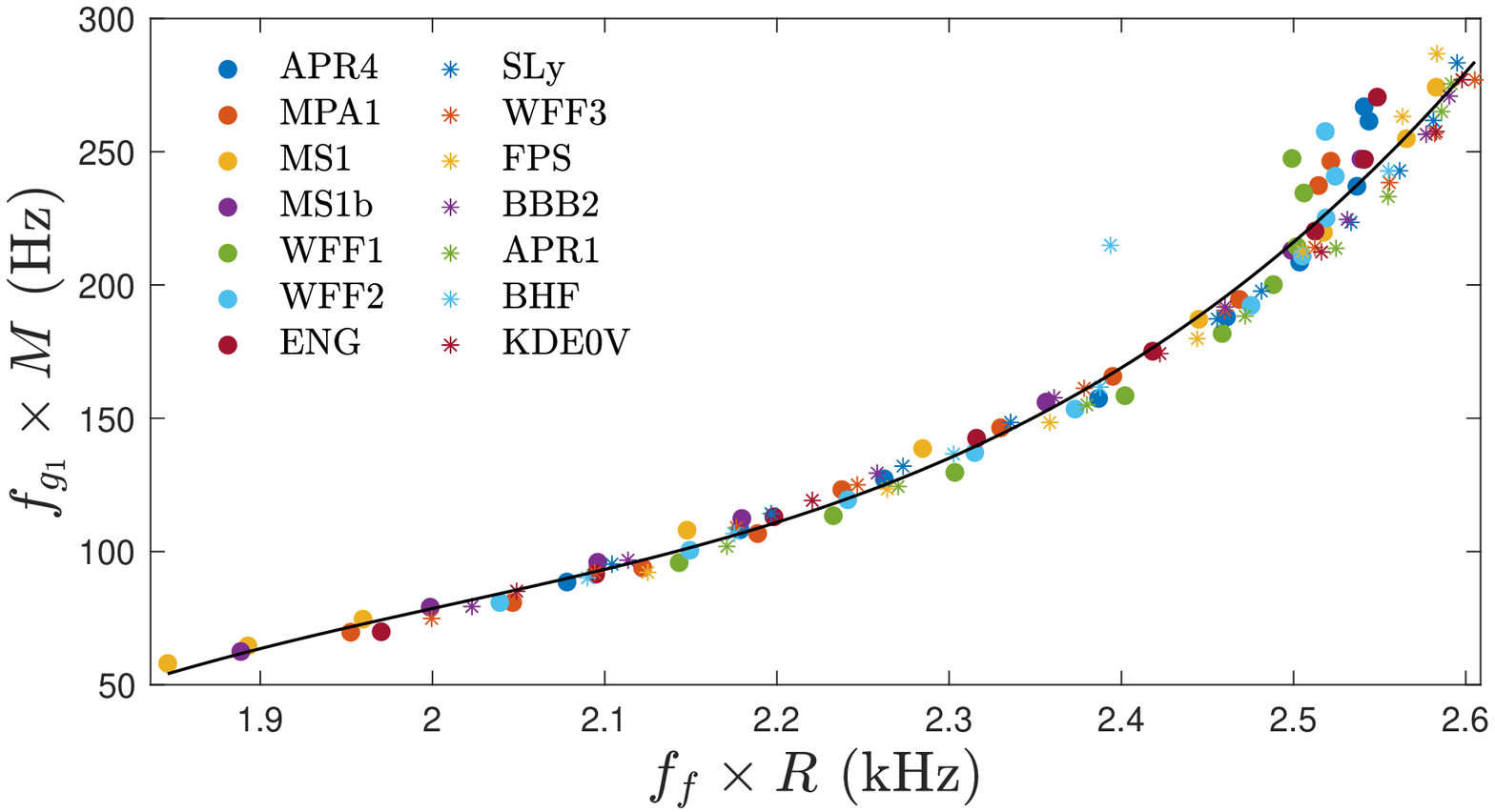}
	\includegraphics[scale=0.43]{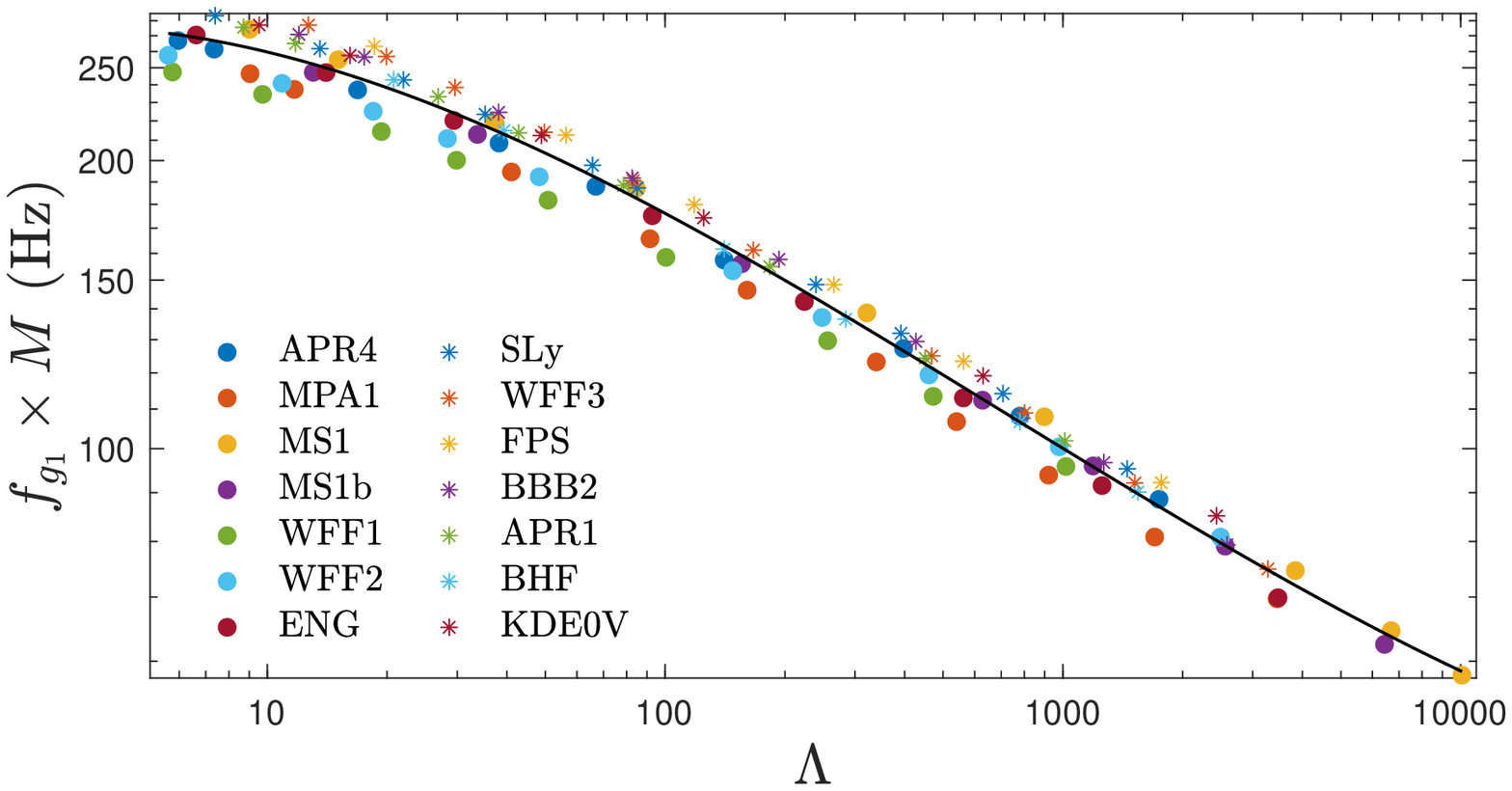}
	\caption{Universal relations for the EOS considered here: the mass-scaled $g_1$-mode frequency as a function of the radius-scaled $f$-mode frequency (top panel), and the compactness-scaled $g_1$-mode frequency as a function of $\Lambda$ (bottom panel).}
	\label{fig:uni}
\end{figure}
\begin{figure}
	\centering
	\includegraphics[scale=0.43]{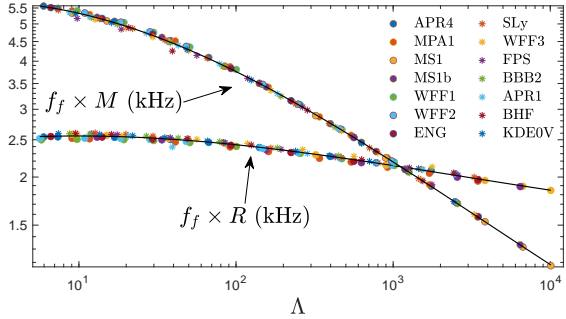}
	\caption{Universal relations established in \protect\cite{Chan14} relating the mass-scaled $f$-mode frequency to $\Lambda$, and relation \eqref{eq:uni3_1} associating the radius-scaled $f$-mode frequency to $\Lambda$, for the EOS considered here.}
	\label{fig:f_uni}
\end{figure}
\begin{figure*}
	\includegraphics[scale=.52]{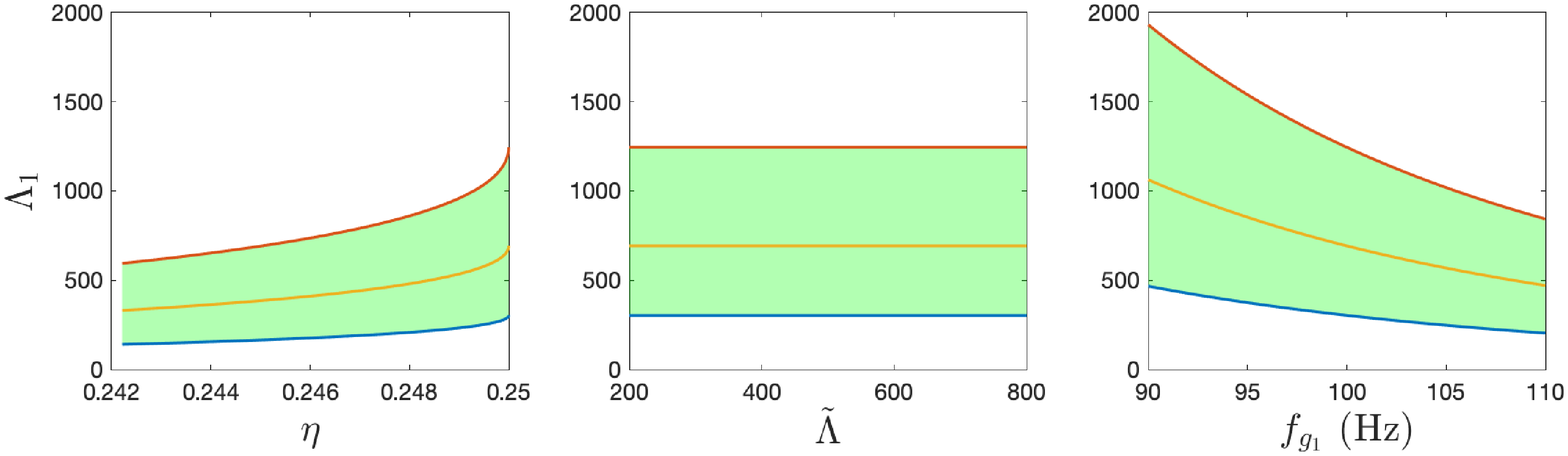}
	\includegraphics[scale=.52]{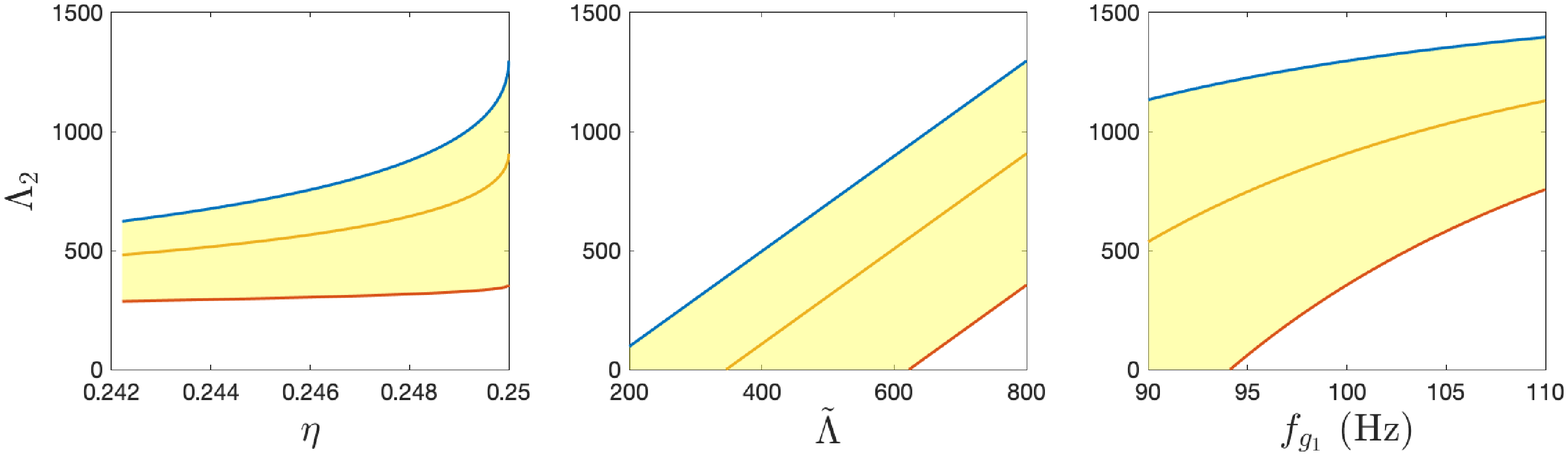}
	\caption{Inferred, individual tidal deformability of the primary (top panel) and the companion (bottom panel), determined via Eqs.~\eqref{eq:uni1}-\eqref{eq:uni3}, using a fixed value of $\mathcal{M}=1.167$. The width of each band is due to an assumed uncertainty in the stratification $\delta$. Three specific $\delta$ are plotted as solid lines on the shaded area: $\delta=0.005$ (blue lines), typical for mature NSs, $\delta=0.01$ (red lines), which may be relevant if the star is heated by some processes (see the main text), and a mid-point stratification $\delta=0.0075$ (orange lines). In each cell, one of $\eta$, $\tilde{\Lambda}$, and $f_{g_1}$ is varied (as per the horizontal axis) while the other two are fixed at ~$\eta=0.25$, $\tilde{\Lambda}=800$, and $f_{g_1}=100$ Hz. 
	}
	\label{fig:rply}
\end{figure*}

Although there are several universal relations in hand, most of them pertain to properties of individual NSs, such as I-Love-Q relation \citep{Yagi13}, and can only become useful when the mass ratio between two components of a binary is provided. We, therefore, are seeking a scheme to obtain the properties of individual star from the observables without the \emph{a priori} input of the mass ratio. We illustrate that with $g$-mode frequency obtained from electromagnetic observations, e.g., the timing of precursors \citep{Kuan21b}, such methodology can be established.

In Fig.~\ref{fig:sheet} we show a universal sheet associating the $g_1$-mode frequency, the tidal deformability, and the stratification. The caveat of this relation is that $\delta$ may be identified with different temperature for different EOS; therefore, it may not be useful even when the temperature can be observed or estimated. To circumvent this dilemma, we can count on an additional information of $g_2$-mode, whose frequency is linearly correlated with that of $g_1$-mode through
\begin{align}
	f_{g_1} = a_0 + a_1f_{g_2},
\end{align}
where the coefficients $a_0$ and $a_1$ are found to be rather insensitive with $\delta$, e.g., $a_0=4.3162$ and $a_1=0.620$ for $\delta=0.005$, and $a_0=5.7221$ and $a_1=0.623$ for $\delta=0.01$. Nonetheless, the difference $f_{g_1}-f_{g_2}$ depends strongly on $\delta$ thus limiting $\delta$ to a certain range if the frequencies of these two modes can be observed. One of the possible scenario, where such detection is plausible, is precursor flares of SGRBs, and we will give a concrete example in Sec.~\ref{sec.IV.A} of limiting the range of $\delta$ via the frequency difference.

Denoting $\delta_5=\delta/0.005$, we plot as the solid line the universal behaviour,
\begin{align}\label{eq:uni1}
	\frac{f_{g_1}}{100 \text{ Hz}} \frac{M}{\sqrt{\delta_5}} =& -48.641+70.782\bigg(\frac{f_f}{\text{1 kHz}}R\bigg)\nonumber\\ &-34.426\bigg(\frac{f_f}{\text{1 kHz}}R\bigg)^2+5.695\bigg(\frac{f_f}{\text{1 kHz}}R\bigg)^3,
\end{align} 
relating the radius-scaled $f$-mode frequency and the mass-scaled $g_1$-mode frequency in the top panel of Fig.~\ref{fig:uni}. The above equation suggests that the quantity of $g_1$-mode frequency divided by $\sqrt{\delta_5}$ depends on the global quantities of the star given $f$-mode frequencies strongly correlated to the mean density, which matches to the indication of Eq.~\eqref{eq:fitomega2del}. Although the relation is valid for, as far as we have checked, $0.001\le\delta\le0.05$, we present only the results for $\delta_5=1$ (coloured dots) in Fig.~\ref{fig:uni} for ease of presentation. In the bottom panel, we plot a certain section of the sheet in Fig.~\ref{fig:sheet}, which shows a bearing on dimensionless tidal deformability via the relation (solid line),
\begin{align}\label{eq:uni2}
	\log\bigg(\frac{f_{g_1}}{100\text{ Hz}} \frac{M}{\sqrt{\delta_5}} \bigg) =&0.411+0.106\log\Lambda-0.126(\log\Lambda)^2 \nonumber\\
	&+0.015(\log\Lambda)^3.
\end{align}
Here the individual quantity $\Lambda$ must not be confused with the observed mutual deformability $\tdm$. Additionally, a relation between the mass-scaled $f$-mode frequency and $\Lambda$ was proposed by \cite{Chan14}, which reads
\begin{align}\label{eq:uni3}
	\log\bigg(\frac{f_f}{\text{kHz}}M\bigg) = 0.814-0.050(\log\Lambda)-0.035(\log\Lambda)^2,
\end{align}
for the EOS considered here. We note that Eqs.~\eqref{eq:uni1} and \eqref{eq:uni2} are invertible maps over the domains of interest, suggesting a one-to-one relation connecting $f_f\times R$ and $\log\Lambda$. Despite the fact that inverting these two equations and then using the common factor $f_{g_1}\times M$ to parametrically fit $f_f\times R$ and $\log\Lambda$ can establish such a relation, we instead directly establish a fitting formula between the data, given by
\begin{align}\label{eq:uni3_1}
    \log\bigg(\frac{f_f}{\text{kHz}}R\bigg) = 0.409+0.013(\log\Lambda)-0.013(\log\Lambda)^2,
\end{align}
which can be viewed as an equivalent of Eq.~\eqref{eq:uni3}. Combining above relations \eqref{eq:uni2} and \eqref{eq:uni3}, the ratio between the frequencies of $f$- and $g_1$-modes can be expressed as a function of $\Lambda$, given by
\begin{align}
	\log\bigg(\frac{f_{g_1}/\sqrt{\delta_5}}{f_f}\bigg)=&-1.403+0.156(\log\Lambda)-0.091(\log\Lambda)^2 \nonumber\\
	&+0.015(\log\Lambda)^3.
	\label{eq:uni4}
\end{align}

The aforementioned relations are powerful in rephrasing observables in terms of unobservable but important quantities. 
Stipulating that we get individual masses from the measurements of the chirp mass and the symmetric mass ratio (see below), and the frequency of $g_1$-mode of the primary from, e.g., precursor observations, we can deduce the individual deformability $\Lambda_1$ from Eq.~\eqref{eq:uni2}. Subsequently, Eq.~\eqref{eq:uni3} translates the mass into $f_f$ of the primary, which then returns the radius $R_1$ via Eq.~\eqref{eq:uni1}. If we also have knowledge of the mutual tidal deformability, the companion's individual deformability $\Lambda_2$ can be obtained which, together its mass, gives the secondary's $f$-mode frequency via Eq.~\eqref{eq:uni3}. Accordingly, Eq.~\eqref{eq:uni4} provides an estimate on the $g_1$-mode frequency of the companion, which, combined with aforementioned companion properties, returns the radius through Eq.~\eqref{eq:uni1}.

In reality, the measurements of $\eta$, $\tdm$, and $\chirp$ all come with error, though the uncertainty on $\chirp$ is typically smaller than the others \citep[see, e.g.,][]{Cutler94,Krolak:1995md}. On top of the above, an uncertainty of $\sim10\text{ Hz}$ in $f_{g_1}$ could also be expected, which effectively translates into error bars for the temperature. All these errors will be reflected in the predictions of Eqs.~\eqref{eq:uni1}-\eqref{eq:uni4}.
As a specific example, we assume $f_{g_1} = 100 \pm 10$~Hz, and adopt the chirp mass $\chirp=1.167$ with $\eta=0.242-0.25$ and the mutual tidal deformability $\tdm=200-800$ of a GW170817-like binary [e.g., \cite{De18,LIGOScientific:2018hze}] in Fig.~\ref{fig:rply}, where the derived tidal deformability of the primary (top panel) and the companion (bottom panel) are plotted for $ 0.005 \leq \delta \leq 0.01$. 
Here the blue lines represent the stratification $\delta=0.005$, typical for mature NSs in coalescing binaries [e.g., \cite{Lai06}], while the red lines depict a relatively large $\delta=0.01$, which may be realised in heated stars (achieved, e.g., through close, tidal interaction or accretion). The mid point, $\delta=0.0075$, is plotted as orange lines. The region bounded by the lines encapsulates the possible range.
We first look at the influence of $\delta$ by taking, for instance, $\eta=0.25$, $\tilde{\Lambda}=800$, and $f_{g_1}=100$ Hz. 
We find for the primary $\Lambda_1=692.72^{+552.47}_{-389.44}$, $R_1=1.55^{+0.42}_{-0.44}$, and $f_f=1829.31^{-256.62}_{+394.23}\text{ Hz}$, while $\Lambda_2=907.28^{-552.47}_{+389.44}$, $R_2=1.26^{-0.20}_{+0.09}$, and $f_f=1708.63^{+454.48}_{-152.89}\text{ Hz}$ for the companion, where the values correspond to $\delta=0.0075$ with the superscript (subscript) associated to $\delta=0.01$ ($\delta=0.005$).

Among the uncertainties in $\chirp$, $\eta$, and $\tdm$, the latter is the most significant since it is extracted from 5-th post-Newtonian (PN) order effects in the gravitational waveform \citep{Flanagan:2007ix,Hinderer09}, which leaves a smaller imprint than the Newtonian-order parameter $\chirp$ and 1PN parameter $\eta$.
Although in our scheme, the large uncertainty in the mutual tidal deformability does not alter the inference of the parameters of the primary if the $g_1$-mode frequency and the masses of the NSs are well-constrained (see, e.g., the second column in the top panel of Fig.~\ref{fig:rply}), 
the inferred properties of the companion are affected considerably. For example, $\Lambda_2=97$--$1297$ and $R_2=0.86$--$1.36$ correspond to this error given $\eta=0.25$, $f_{g_1}=100\text{ Hz}$, $\delta=0.005$, and the above parameters of the primary. Physically speaking, this is because we attribute a $g$-mode to the primary and not the secondary, so that less information is obtained about the latter using the formulae derived here. In addition, the predictions are sensitive to $\eta$ and $f_{g_1}$ as illustrated in Fig.~\ref{fig:rply}, where the inferred tidal deformability can drop by $50\%$ from symmetric case ($\eta=0.25$) to a mildly asymmetric case with $\eta=0.242$. On the other hand, they can increase by a factor of two if the $g_1$-mode frequency of the primary is $f_{g_1}=90$ Hz instead of 110 Hz. 
 
Although the individual masses $M_1$ and $M_2$ can be determined by the chirp mass and the symmetric mass ratio measurements, with an error mainly due to the latter [see, e.g., Sec.~5 and 6 of \cite{Kokkotas94} for a discussion], the radii of constituents can not be easily constrained. The $g$-mode methodology, however, can additionally determine the radii of binary members with high accuracy if the temperature is modest. As the radius of NSs with a mass of $1.4M_{\odot}$ still spans a range of $11$--$13$ km for EOS candidates that pass the observations [e.g., \cite{Capano:2019eae,Landry:2020vaw,Dietrich:2020efo,Pang:2021jta}], such a novel way to measure the radius may therefore constitute a valuable tool in whittling down the pool of currently-viable EOS.

On top of this arguably already executable application of universal relations, we speculate on the possibility of detecting, though indirectly, the $f$-mode frequency from the accumulation of phase, $\delta\Phi_{\text{dyn}}$. This is expected to be a plausible with detectors in the near future. In particular, $f_f$ of the progenitors of GW 170817 may be determined to within tens of Hz with the Einstein Telescope \citep{Pratten19}. 

\begin{figure}
	\centering\hspace*{-.6cm}
	\includegraphics[scale=.4]{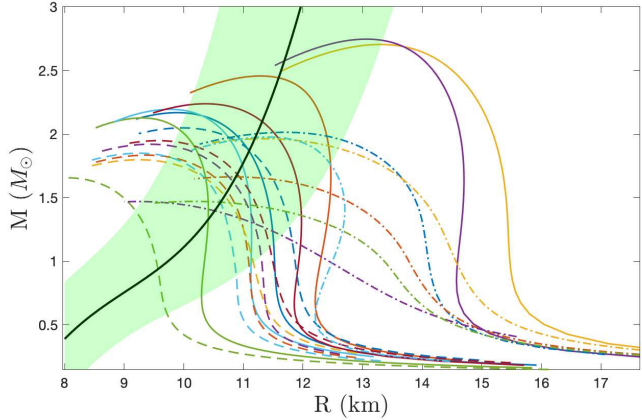}
	\caption{The curves same in Fig.~\ref{fig:EOS} overlapped with Eq.~\eqref{eq:uni1}. The solid line stands for the case with $f_{f}=2.2 \text{ kHz}$ and $f_{g_1}=100 \text{ Hz}$, and the shaded area corresponds to the range of $f_{f}=2-2.4 \text{ kHz}$ and $f_{g_1}=80-120 \text{ Hz}$.}
	\label{fig:uni1}
\end{figure}

The determination of $f_f$ has a double-dose of implications. Firstly, incorporation of Eqs.~\eqref{eq:uni1}-\eqref{eq:uni3} results in an ``on-shell'' (physically realisable) condition of $M(R)$ if $f_f$ and $f_{g_1}$ are provided. In Fig.~\ref{fig:uni1}, we overlap the ``on-shell'' $M(R)$ to the mass-to-radius diagram of the EOS considered here for the typical values of $f_f=2-2.2$ kHz and $f_g=80-120$ Hz. Here the solid line corresponding to $f_f=2$ kHz and $f_{g_1}=100 \text{ Hz}$. Secondly, we can know simultaneously the mean density due to its universal relation with $f$-mode frequency \citep{Chan14,Kruger19}. According to the derived mean density and the $g_1$-mode frequency, we can represent the NS on Fig.~\ref{fig:grouping}, and thereby sift the branch of EOS for this system. We note that even for binaries without precursors, $f_{g_1}$ can be determined through Eq.~\eqref{eq:uni2} if the tidal deformability $\Lambda$ is measured.

Another scenario where these EOS-insensitive expressions are particularly helpful is black hole-NS binaries since the individual tidal deformability equals to the mutual one, which is an observable, thus activating the Eqs.~\eqref{eq:uni2}-\eqref{eq:uni4} without the additional processing translating observables to $\Lambda$.

%%%%%%%%%%%%%%%%%%%%%%%%%%%%%%%%
%%%%%%%%%%%%%%%%%%%%%%%%%%%%%%%%

%%%%%%%%%%%%%%%%%%%%%%%%%%%%%%%%
\subsection{Speed of sound}
\label{sec.III.B}
%%%%%%%%%%%%%%%%%%%%%%%%%%%%%%%%
%%%%%%%%%%%%%%%%%%%%%%%%%%%%%%%%
\begin{figure}
	\centering
	\includegraphics[scale=0.43]{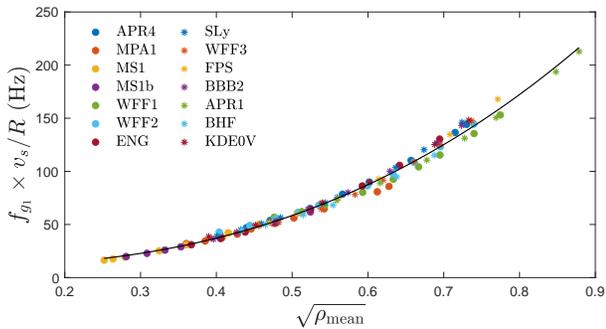}
	\caption{Rescaled $g_1$-mode frequency as a function of $\md$ for the EOS considered here, with $v_s$ being the central sound speed.}
	\label{fig:sound}
\end{figure}
%%%%%%%%%%%%%%%%%%%%%%%%%%%%%%%%

The measurement of $g_1$-mode can be translated into the central sound speed of the equilibrium via the strong correlation,
	\begin{align}\label{eq:uni5}
		\frac{f_{g_1}/\sqrt{\delta_5}}{100 \text{ Hz}}\bigg(\frac{v_s}{R}\bigg) = &0.124 - 0.296 \md^{0.5} + 1.685 \md \nonumber\\&+ 1.442 \md^{1.5},
	\end{align}
	shown as the solid black line in Fig.~\ref{fig:sound}. Given that the stiffness of a EOS can be characterised by the maximum sound speed \citep{VanOeveren17}, this central value may be informative enough for shedding light on the stiffness of the EOS since the sound speed culminates at the center. The stiffness of EOS affects the bulk properties of NSs, such as the maximum mass that is supportable with the EOS \citep{Moustakidis16,Zhang19}, and the tidal deformability \citep{VanOeveren17,Kanakis-Pegios20}; therefore, the constraint on the stiffness may be translated to either the upper or the lower bound for the sound speed. For instance, the lower bound of sound speed should be $\sqrt{0.6}$ if the secondary of the binary hosting GW 190814 is a NS \citep{Tews20}.  In light of the constraints set by GW 170817, the determination of the stiffness can augment the GW channel to limit viable EOS. Another aspect of utilising the speed of sound to benefit astrophysics can be found in, e.g., \cite{Tews18}, where the authors build a family of nuclear interactions in terms of the behaviour of $v_s$, thus the inferred $v_s$ from Eq.~\eqref{eq:uni4} limits the EOS model.

%%%%%%%%%%%%%%%%%%%%%%%%%%%%%%%%
\section{g-mode candidate systems}
\label{sec.IV}
%%%%%%%%%%%%%%%%%%%%%%%%%%%%%%%%
To better utilise the universal relations involving $g$-modes developed in the present article, we propose some candidate systems where $g$-mode detection may be plausible.

\subsection{Precursor flares}
\label{sec.IV.A}
For the NS progenitor of a merger, it has been proposed that mode resonance(s) may trigger precursor flashes of the SGRB following the merger, notably via interface- \citep{Tsang11,Tsang13} and $g$-modes \citep{Kuan21b} for slow-rotating NSs, and $f$- and/or $r$-modes for fast-spinning NSs \citep{Suvorov:2020tmk}. The uncertainties in both the spin rate of NS progenitors and the jet formation timescale of the main episode allow for candidates of several kinds of mode, where the former quantity shifts the mode frequency relative to the orbital frequency inferred from the occurrence of precursors, and the latter timescale blurs the amount of time prior to the merger\footnote{Precursors can only be timed relative to the main burst by definition, which occurs only some time after the merger since jet formation is not instantaneous, but has a development timescale that depends on its formation mechanism [see Sec.~II.~A of \cite{Suvorov:2020tmk} for a discussion].}.

Seeing that NSs in coalescing binaries are expected to be mature and thus slowly rotating, we examine in a separate work \citep{prep} how two precursor flares \cite[associated, for example, with GRB 090510][] {Troja:2010zm} can be accommodated by the lowest two orders of either $g$-modes or interface-modes. Nevertheless, the indicated spin rate of the host NS is $\lesssim50$ Hz for the former modes (see also below), while $\ll10 \text{ Hz}$ is suggested by the latter modes. Even so, however, we emphasise here that we restrict ourselves to the $g$-mode resonance scenario \citep{Kuan21b}, but other (e.g., $i$-mode) possibilities exist \citep{Tsang11,Tsang13}. The timing of precursors thus allows one to infer the (inertial-frame) frequency of the relevant mode, given by
\begin{align}
	\omega \simeq \omega_0 - 0.89 m\rot 
\end{align}
when the stellar magnetic field strength is small ($B\lesssim 10^{14}$ G), where $\omega_0$ is the free mode frequency (i.e., the frequency not accounting for tidal modulations to stellar structure),  $\rot$ is the spin rate of the star, and $m$ is the winding number of the mode [cf.~Eq.~(70) in \cite{Kuan21a}]. The numerical factor $0.89$ arises by considering leading-order corrections in the slow-rotation approximation, though is also, in principle, sensitive to the compactness of the star [see Fig.~(10) in \cite{Kuan21a}]. 

For the specific case of SGRB 090510, the two precursors are detected when $\Oo\approx160$ Hz and $\Oo\approx510$ Hz, respectively. If they are attributed to $l=m=2$ $g_1$- and $g_2$-modes of the star, we obtain the equalities
\begin{subequations}
\begin{align}\label{eq:precusor1}
	\omega_{g_1} - 1.78 \rot = 510 \text{ Hz},
\end{align}
and
\begin{align}\label{eq:precusor2}
	\omega_{g_2} - 1.78 \rot = 160 \text{ Hz},
\end{align}
\end{subequations}
associating with the frequencies of the $g_{1}$- and $g_2$-modes of the primary. The notation $\omega=2\pi f$ gives the angular frequency for linear frequency $f$. Exploiting the fact that the ratio between the frequencies of $g_1$- and $g_2$-modes is roughly a constant, viz.~$\omega_{g_2}/\omega_{g_1} \simeq 0.68$ [Eq.~\eqref{eq:ratio12}], regardless the EOS and $\delta$, we deduce that $\omega_{g_1}=174\times 2\pi$ Hz, $\omega_{g_2}=118\times 2\pi$ Hz, and $\rot=52.2\times 2\pi$ Hz. The $g_1$-mode frequency can set a lower bound for $\delta$ since $\omega_1$ is smaller for decreasing $\delta$. In addition, although the ratio $\omega_{g_2}/\omega_{g_1}$ varies only slightly with respect to EOS and $\delta$, the difference $\omega_{g_1}-\omega_{g_2}$ requires $\delta$ to fall in certain range resulted from its dependence on $\delta$, as well as EOS. For instance, the difference of $\omega_{g_1}-\omega_{g_2}=55.7\times 2\pi$ Hz here restrains the stratification to be $0.008\lesssim\delta\lesssim 0.2$ for EOS APR4, and $0.006\lesssim\delta\lesssim 0.44$ for EOS ENG. 

We should emphasise that aforementioned requirements for $\omega_{g_1}$, $\omega_{g_2}$, and $\delta$ are the necessary conditions for these two $g$-modes to be responsible for precursors; the sufficient condition is that the mode amplitude can be resonantly excited beyond certain threshold value \citep{Kuan21a,Kuan21b}. It turns out the necessary and sufficient requirements are stringent for stellar parameters, and the viable region on the parameter space is expected to be narrow [see \cite{Kuan21b} for a thorough discussion]. Nonetheless, we note that the EOS of Group I are favoured in igniting precursors with moderately massive or light stars. Although the $g$-mode resonance in the quite massive ($\gtrsim 2M_{\odot}$) or light ($\sim 1M_{\odot}$) stars for the EOS of Group II are possible to fuel these early flares, it is unlikely that stars with EOS of Group III can host such pre-emissions.

In addition, the inferred $g_1$-mode is close to $200$ Hz, which is marginally high in terms of its typical value $\sim 100$ Hz. The tidal heating resulted from $g$- and $f$-modes investigated in Sec.~\ref{sec.II.B} may pose a tension between the realisability of such frequency. Nonetheless, it is possible for a spinning NS to have low amplitude $r$-mode excitations, which may not slow down the star but still heat up the material to make the NS hot enough to admit such $g$-modes.

\subsection{Quasi-periodic oscillations}
\label{sec.IV.B}

Apart from the possibility of precursor flashes observed prior to merger events, there have not yet been direct $g$-mode observations in neutron stars, either of the nascent or mature variety. It has, however, been speculated by \cite{Strohmayer13} that $g$-modes may be responsible for the QPOs observed in the X-ray light curves of accreting millisecond X-ray pulsars (AMXPs).
While it is unclear which particular modes may be responsible in these cases (and indeed it is usually thought that \emph{disk} g-modes may be responsible rather than \emph{stellar} g-modes), it is necessary that their (inertial-frame) frequency matches that was observed, which is theoretically possible for the class of $g$-modes considered here. 

We take XTE J1751-305 and XTE J1807-294 as examples since their QPO frequencies lie in the range of interest (i.e., $\lesssim 400$ Hz).
Although there are other possibilities, $r$- and $g$-modes are candidates for matching to these frequencies \citep{Strohmayer13,Andersson14,Lee14}. The excitation of the former modes would generally be expected to sap angular momentum from the star also, especially if they are driven CFS unstable. However, at least for the specific outburst observed in XTE J1751-305, it is likely that if $r$-modes were responsible, the spin-up behaviour of the star following the first X-ray pulse in 2002 (the small variation in the flux is only observed in this stage) would have been lower \citep{Andersson14}, which brings $g$-modes to our attention. Nevertheless, we emphasise that the trigger mechanism for QPOs is not well-established, and we consider $g$-modes in this context to demonstrate how one may phenomenologically pin the pulsar in the $g_1$-space given an observation. The above-described scenario is represented by point (D) in Fig.~\ref{fig:g_space}.

According to Eq.~\eqref{eq:omega2T}, the mass, the radius, and the temperature of stars must be known in order to mark them on the $g$-space. Although we can deduce the mean density of two AMXPs, viz.~the range of $0.919$-$1.214$ for\footnote{Note that this density is inferred from the calculations in \cite{Andersson14}, who operated under the assumption that an $r$-mode was responsible for the outburst event, and it is therefore, strictly speaking, inconsistent to use their values in a $g$-mode analysis here. We adopt these values however to offer a proof-of-principle.} XTE J1751-305 \citep{Andersson14}, and the range of $0.499$-$1.425$ for XTE J1807-294 \citep{Leahy11}, the lack of the knowledge of their temperature prevent us from locating them on \emph{$g$-space}. To better utilise the tool of $g$-space, future measurement of the temperature is crucial.

\section{Discussion}\label{sec.V}
In this article we investigate the dependence of $g$-mode frequencies on both microscopic (e.g., local changes in the adiabatic index) and macroscopic (e.g., stellar compactness) physics, revealing three families of EOS which still somehow support a number of universal relations. In particular, we found that $g$-mode frequencies correlate to the temperature-modified mean density to the power of a parameter $\varsigma$ [Eq.~\eqref{eq:omega2T}], where the range of $\varsigma$ divides EOS considered here (Tab.~\ref{tab:constant}). This empirical fitting formulae define the \emph{$g$-space}, where a given point corresponds to a unique EOS, which is split into three bands depending on $\varsigma$. Given some observations, from, e.g., precursors and QPOs, points can be pinned down in the \emph{$g$-space} to scrutinise the EOS candidate (Fig.~\ref{fig:g_space}).
Although joint constraints have been studied via Bayesian analysis in, e.g., \cite{Miller19,Raaijmakers19,Raaijmakers21}, our approach here is not interlaced with uncertainties in the prior of the parameterisation of the EOS. In addition, the appliance of Eq.~\eqref{eq:omega2T} to $g_1$- to $g_3$-modes implies that the ratios between any two of them are roughly constant. Furthermore, if we are able to measure both $g_1$- and $g_2$-modes in a NS, e.g., from two precursors in a SGRB [the double events in GRB090510 may serve as an example \citep{Kuan21b}], we can measure the spin of the star within some tolerance by solving Eqs.~\eqref{eq:precusor1} and \eqref{eq:precusor2}. 

On top of the trifurcation, universal dependences of $g$-modes can be found in terms of certain quantities [Eqs.~\eqref{eq:uni1}-\eqref{eq:uni3}]. These EOS insensitive information can aid in placing constraints on EOS without \emph{a priori} knowledge of the EOS. We show the universal sheet among the $g_1$-mode frequency, the tidal deformability, the stratification, and the mass of NSs (Fig.~\ref{fig:sheet}). Although inferring $\delta$ in a certain system is non-trivial, some estimates will be possible if we can measure additionally $f_{g_2}$. For a particular value of $\delta$, a specific section of the universal sheet will be picked up; assuming a typical value $\delta=0.005$, we reduce the sheet to Eq.~\eqref{eq:uni1}. For this $\delta$, we also establish the relation between $f_{g_1}$, $M$, and $\Lambda$ [Eq.~\eqref{eq:uni2}]. Three prospective applications of the above universal relations can be summarised as:
\begin{enumerate}
	\item Assuming we can measure $\mathcal{M}$, $\eta$, $\tdm$, and $f_{g_1}$: the mass $M$, the radius $R$, and the $f$-mode frequency $f_f$ can be inferred from Eqs.~\eqref{eq:uni1}-\eqref{eq:uni3}. The particular case of a GW170817-like binary is discussed in Sec.~\ref{sec.III.A}.
	\item Assuming we can measure the $f$- and $g_1$-modes: a region of feasible models can be drawn on the mass-to-radius diagram (Fig.~\ref{fig:uni1}); such measurements are expected to be plausible with near-future detectors.
	\item Given $\Lambda$ and the $f$-mode frequency: the $g_1$-mode value $f_{g_1}$ can be extracted via Eq.~\eqref{eq:uni4}. In addition, the strong correlation between $f_f$ and $\md$ gives an estimate of the latter quantity, which delegates $f_{g_1}$ on a certain region in Fig.~\ref{fig:grouping}, thus picking a certain group of EOS for the system.
\end{enumerate}

Besides the global quantities, Eq.~\eqref{eq:uni5} sheds light on the central sound speed if the radius and the $g_1$-mode frequency are available. Seeing that the central sound speed of the NSs near the EOS-related maximal mass reflects the ``stiffness'' of EOS, this indicates another aspect to disfavour certain range of ``stiffness'' other than the analysis having been done to GW 170817, which is based on the measurement of $\tdm$ \citep{LIGOScientific18}.

In addition, universal relations may also be leveraged to pin down observations on $g$-space if we can maneuver out of the observables the mass, radius, and temperature of NSs. As a na{\"i}ve example, we consider the remnant of merger. The transcendental information associating the tidal deformability of the \emph{progenitors} of mergers to the compactness of the long-lived \emph{remnant} [cf.~Eq.~(4) and Fig.~9 of \cite{Manoharan21}] says that we can learn the compactness of the remnant $\cpt_{\text{rem}}$ from the measurement of $\tdm$. In addition, the $f$-mode frequency of the remnant, $f_{f,\text{rem}}$, may be measured from its influence on SGRB photon counts following the merger since the jet opening angle may widen and shorten as the star oscillates \citep{Chirenti19}.
Envisaging we have $f_{f,\text{rem}}$ and $\cpt_{\text{rem}}$, the five stellar parameters $f_{g_{1},\text{rem}}$, $f_{f,\text{rem}}$, $M_{\text{rem}}$, $R_{\text{rem}}$, and $\Lambda_{\text{rem}}$ of the remnant can be calculated by mutually solving Eqs.~\eqref{eq:uni1}-\eqref{eq:uni3}. This information suffices to pin the system down in $g$-space within some tolerance.

To make further use of $g$-mode asteroseismology via the universal relations, we point out two possible avenues for detecting $g$-modes via electromagnetic sector, viz.~the pre-emissions of SGRBs (Sec.~\ref{sec.IV.A}) and (more speculatively) the QPOs in the X-ray light curve of AMXPs (Sec.~\ref{sec.IV.B}). If the temperature in these systems can be measured somehow with future observatories (e.g., phase-resolved hot spot tracking for the latter), including the merger remnants, the $g$-mode ansatz introduced in this work, as well as the universal relations provided here, can be readily applied to place constraints on the EOS by identifying observations on \emph{$g$-space}, as demonstrated in Fig.~\ref{fig:g_space}.

\section*{Acknowledgement}
CK gratefully acknowledges financial support by DFG research Grant No. 413873357. HJK recognises support from Sandwich grant (JYP) No. 109-2927-I- 007-503 by DAAD and MOST during early stages of this work. We are grateful for the useful comments/suggestions given by the anonymous referee, which improved the quality of the present article.

\section*{Data availability statement}
Observational data used in this paper are quoted from the cited works. Data generated from computations are reported in the body of the paper. Additional data will be made available upon reasonable request.

%%%%%%%%%%%%%%%%%%%%%%%%%%%%%%%%%%%%%%%%

\bibliographystyle{mnras}
\bibliography{references}

\appendix

\section{Tidally-neutral $g$-modes}
\label{appendix}
\begin{figure}
	\centering
	\includegraphics[scale=0.4]{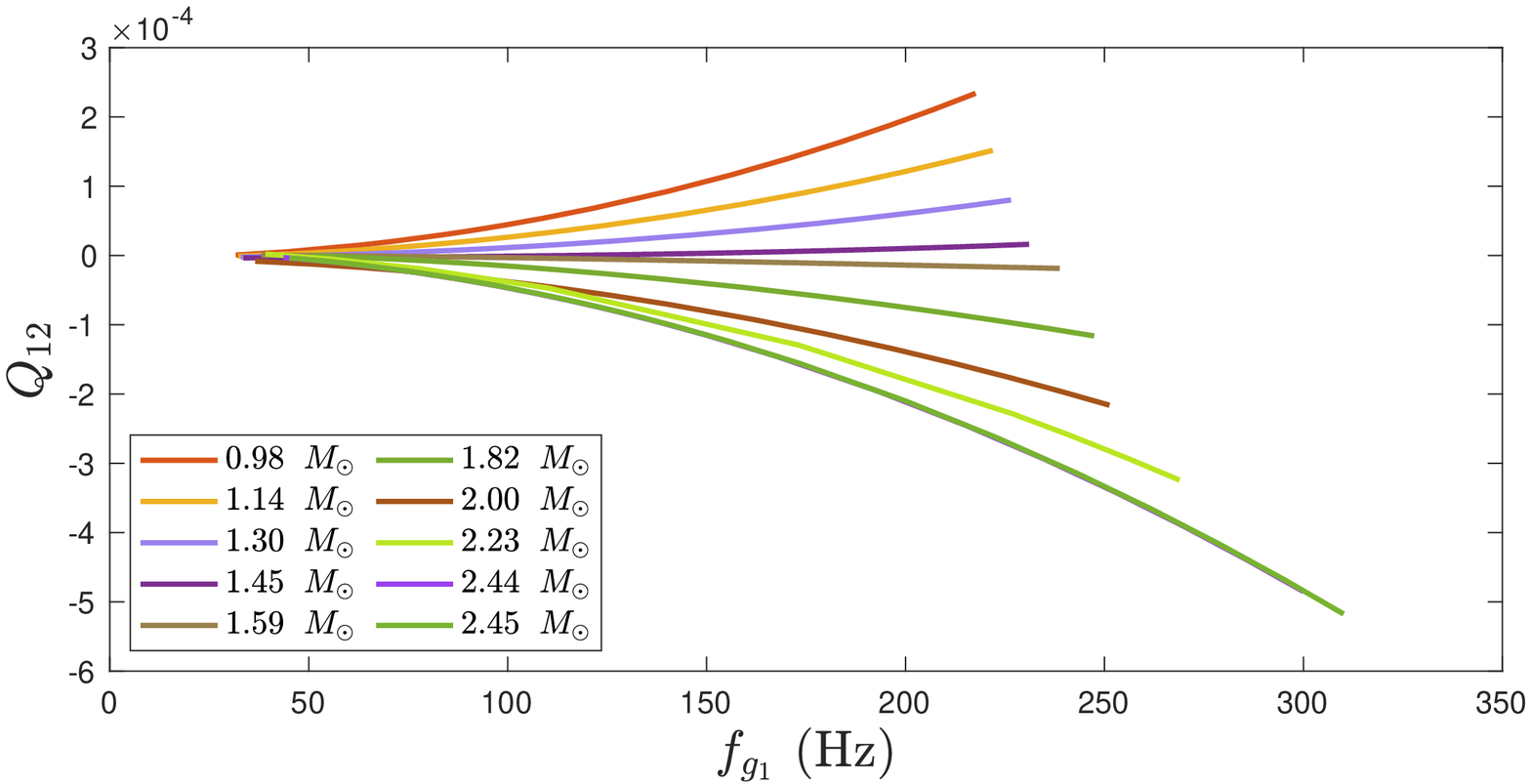}
	\includegraphics[scale=0.4]{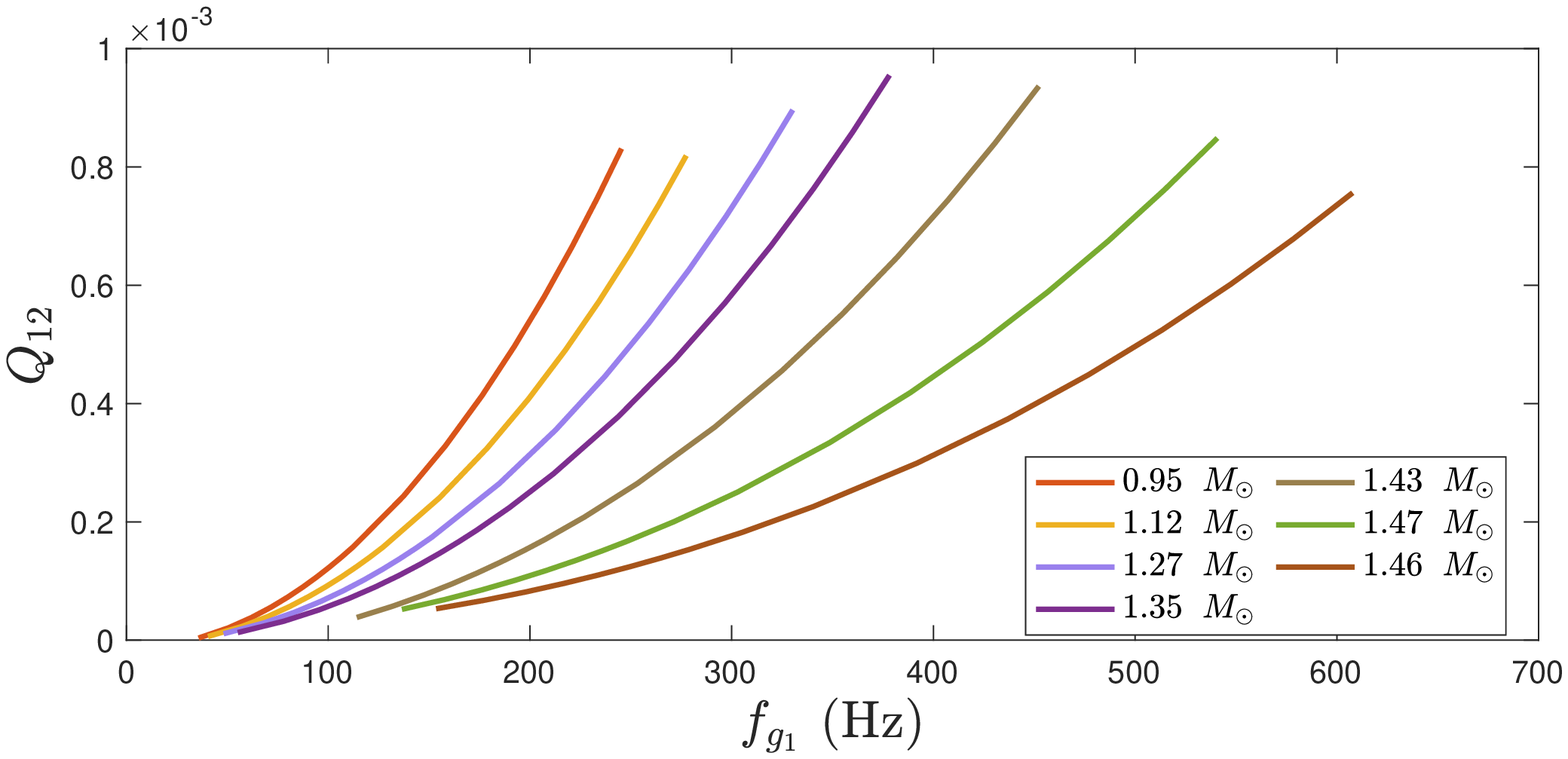}
	\caption{Tidal coupling strength as a function of $g_1$-mode frequency for several models of EOS MPA1 (top panel) and PCL2 (bottom panel). The mass of the star is held fixed for each curve shown here, which is drawn by increasing the stratification index $\delta$.}
	\label{fig:over}
\end{figure}
In binaries, QNMs will be driven by the exertion of tidal forces from the companion. The resulting motions generate excess in gravitational energy by absorbing and localising the orbital energy into the star. The ``efficiency'' of the energy-soaking process for a specific mode is appraised by the overlap between the associated motion $\xi_{lnm}$ and the gravitational potential $\Phi^T$ of the companion \citep{Press77,Kuan21a}, where $\{nlm\}$ are conventional quantum numbers for harmonic oscillators. Considering solely the quadrupolar component ($l=2$) of the potential, $\Phi^{T}_{qd}$, due to its predominant contribution to tidal effects [see, e.g., the discussion in the Appendix A.2 of \cite{Damour12}], tidal coupling strength of QNMs to the (leading-order) external tidal field is given by \cite{Kuan21a}
\begin{align}\label{eq:over}
	Q_{nl} = \frac{1}{MR^{l}} \int d\mathcal{V}\alpha(p+\rho) \bar{\xi}_{nll}^{\mu} \nabla_{\mu} (\Phi^{T}_{\text{qd}}),
\end{align}
which is independent of the winding number $m$. Here $d\mathcal{V}$ is the spatial volume form, $\alpha$ is the red-shift factor, while $p$ denotes the pressure profile of the star. Equation \eqref{eq:over} indicates that only $l=2$ QNMs are relevant to the leading-order tidal phenomena due to the orthogonality between spherical harmonic functions $Y_{lm}$.

Although only the absolute value of the tidal coupling $Q_{nl}$ matters in the context of energy absorption, we find that the $n=1, l=2$ $g$-mode (quadrupolar $g_1$-mode) may have positive and negative values of $Q_{12}$, implying that there are some $g_1$-modes with $Q_{12}=0$ from continuity (Rolle's theorem). For instance, we plot the tidal coupling of $g_1$-modes for several masses with EOS MPA1 in the top panel of Fig.~\ref{fig:over}, where each curve shows the $g_1$-modes with different stratification. We see that $g_1$-modes in the star with $1.59M_{\odot}$ have nearly vanishing $Q_{12}$, and may thus be fitting called `tidally-neutral'. By contrast, we find that $Q_{12}$ is always positive for the EOS belonging to \emph{Group III} (see, e.g., EOS PCL2 in the bottom panel of Fig.~\ref{fig:over}) except for EOS H7 and GNH3, whose models with mass close to the maximum have $Q_{12}\simeq 0$.
\label{lastpage}

\end{document}